\documentclass[11pt,twoside]{article}

\usepackage{asp2004}
\usepackage{epsf}
\usepackage{lscape}

\begin{document}
\title{On the Gravitational Boundedness of Small Scale Structures in Molecular Clouds}   
\author{Sami Dib and Jongsoo Kim}   
\affil{Korea Astronomy and Space Science Institute, 61-1, Hwaam-dong, Yuseong-gu, Daejeon 305-348, Korea; dib@kasi.re.kr; jskim@kasi.re.kr}    

\begin{abstract} 
 We investigate, in a set of 3D numerical simulations of driven, magnetized, isothermal, and self-gravitating molecular clouds (MCs), the statistical correlations between the energy ratios (thermal/gravity, and kinetic/gravity) of clumps and cores (CCs) identified in the simulations and gravitational binding indicators commonly used in observational studies such as the Jeans number, $J_{c}$, and the virial parameter, $\alpha_{vir}$. In the energy ratios, we consider the surface energy terms which account for the effects of the environment on the clump gravitational boundedness. We find that: a) $J_{c}$ and the thermal/gravitational energy ratios are well correlated, b) $\alpha_{vir}$ and the (thermal+kinetic)/gravity or kinetic/gravity energy ratios are poorly correlated, additionally affected by the ambiguity of the compressive or dispersive effect of the velocity field. This result suggest that the use of $\alpha_{vir}$ estimates in the observations is only useful to assess the kinetic+thermal energy content of a CC and not its gravitational boundedness. Finally, we discuss briefly the possibility of measuring the kinetic energy surface term directly in the observations.
\end{abstract}

\section{Introduction}   
\begin{figure}
\plottwo{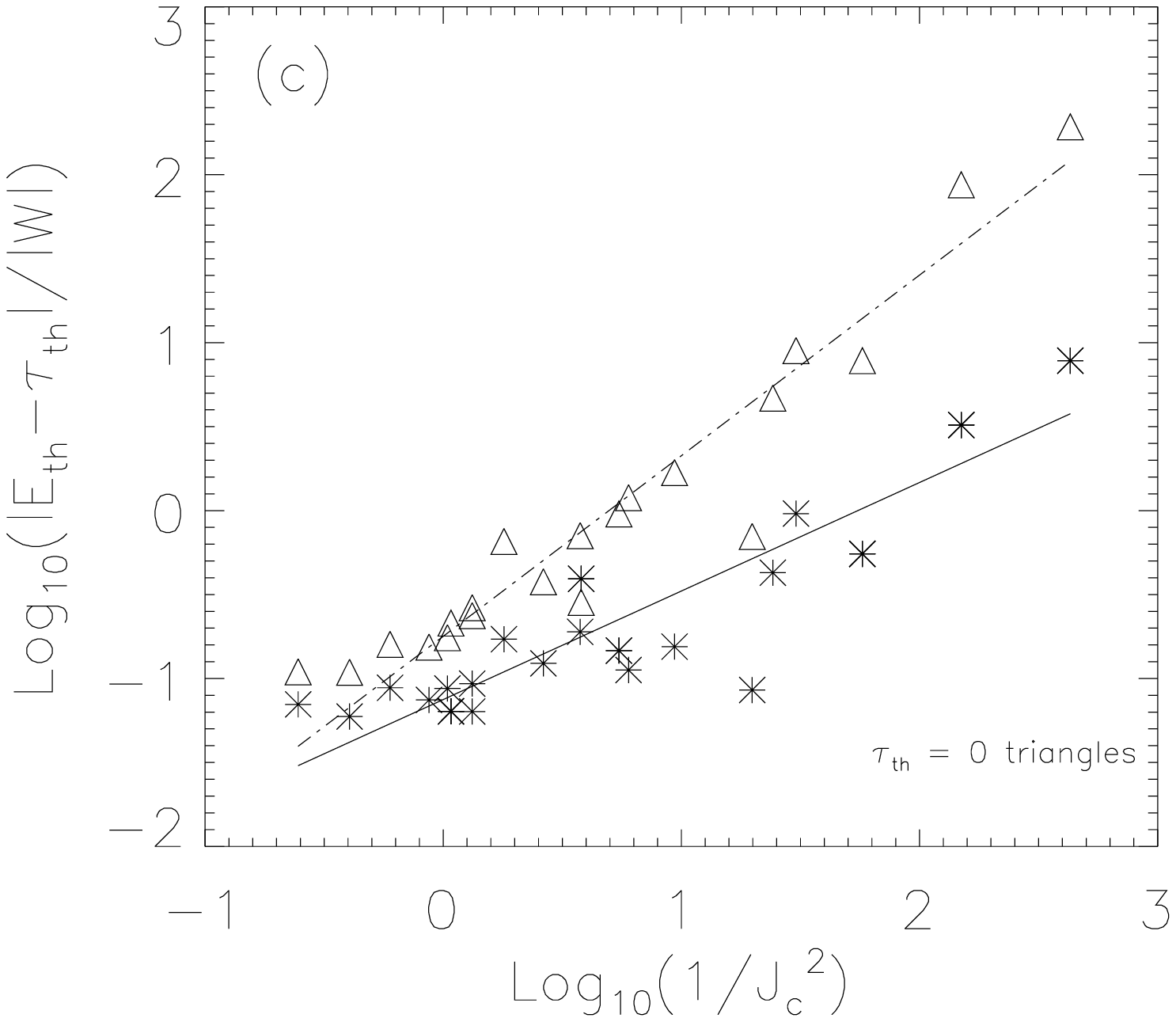} {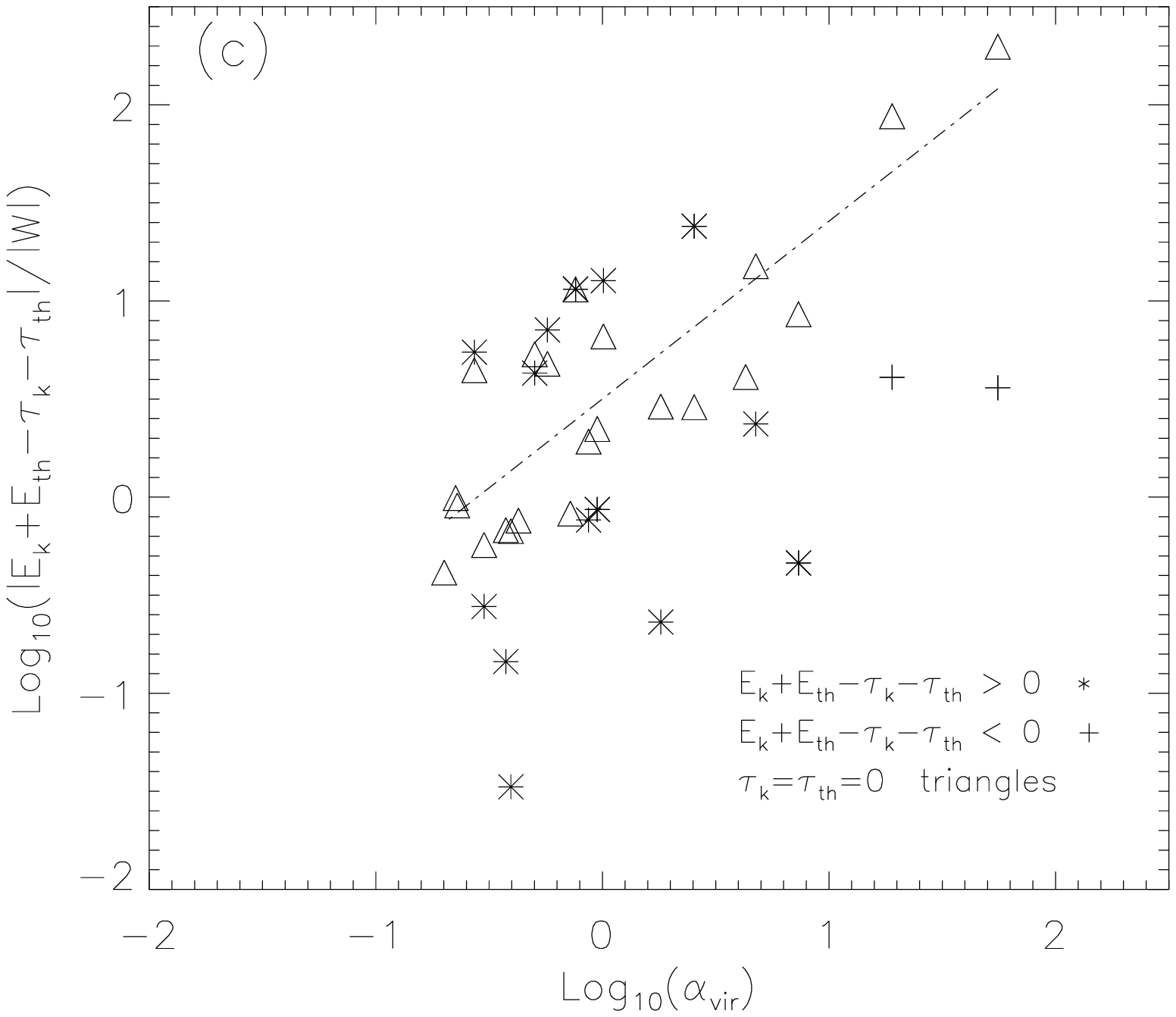}
\caption{ (Left) Relationship between $J_{c}$ and the ratio of thermal to gravitational energies for CCs in a strongly supercritical cloud model. All clumps have $(E_{th}-\tau_{th}) > 0$. The full line is a fit to the data when the thermal surface energy terms are taken into account (stars) and the dot-dashed line when they are not (triangles). (Right) Relationship between $\alpha_{vir}$ and the ratio of the kinetic+thermal to gravitational energies in the same model. CCs are cataloged by whether kinetic+thermal energies provide a net support to the clump ($E_{k}+E_{th}-\tau_{k}-\tau_{th} > 0$, stars), or a net compressive effect ($E_{k}+E_{th}-\tau_{k}-\tau_{th} < 0$, crosses), or whether the surface energy terms are omitted (triangles).}
\label{fig1}
\end{figure}

One of the most challenging issues in the field of star formation is to assess the gravitational boundedness of clumps and cores (CCs) which are prone to star formation. Constraining the masses and scales on which CCs are gravitationally bound sets important constraints on the star formation efficiency in molecular clouds (MCs). Two quantities commonly used in the observations in order to assess the gravitational boundedness of CCs are the Jeans number $J_{c}$ and the virial parameter $\alpha_{vir}$. $J_{c}$ measures the importance of the thermal support against gravity and is defined as: $J_{c}=R_{c}/L_{J,c}$, where $L_{J,c}=(\pi c_{s}^{2}/G \bar{\rho_{c}})^{1/2}$ is the mean Jeans length in the core, $\bar{\rho_{c}}$ the average density, and $c_{s}$ the thermal sound speed, and $R_{c}$ the characteristic size. The virial parameter measures the ratio of thermal+kinetic to gravitational energies (e.g., Bertoldi \& McKee 1992) and is defined as: $\alpha_{vir}= (5~\sigma^{2}~R_{c}/G~M_{c})$, where $M_{c}$ and $\sigma$ are the mass and one dimensional density-weighted velocity dispersion of the CC. 

Another method for evaluating the gravitational boundedness of CCs is by using the virial theorem (VT). As shown in Dib et al. (2006a), the surface energy terms in the VT are equally important for the overall virial balance. Thus, in order to compare the gravitational term $W$ in the VT to the net effect of other forces i.e., $|2(E_{th}+E_{k}-\tau_{th}-\tau_{k})+E_{m}+\tau_{m})|$\footnote{$E_{th}=\frac{3}{2}\int_{V} P dV$ is the volume thermal energy, where $P$ is the thermal pressure, $E_{k}=\frac{1}{2}\int_{V} m_{i} v_{i}^{2} dV$ the volume kinetic energy, $E_{m}=\frac{1}{8 \pi}\int_{V} B^{2} dV$ the volume magnetic energy, $\tau_{th}=\frac{1}{2}\oint_{S} r_{i} P \hat{n}_{i} dS$ the surface thermal energy, $\tau_{k}=\frac{1}{2} \oint_{S} r_{i} \rho v_{i} v_{j} \hat{n}_{j} dS$ the surface kinetic energy, $\tau_{m}=\oint_{S} r_{i} T_{ij} \hat{n}_{j} dS$ the surface magnetic energy, where $T_{ij}$ is the Maxwell stress tensor given by $T = (1/4~\pi)~({\bf B~B}-0.5~B^{2}{\bf I})$ and $I$ the identity tensor, ${\bf r}$ is the position of each particle/cell with respect to the center of mass (CM), and ${\bf v}$, the relative velocity of particles/cells to the CM. The gravitational term is defined as $W=-\int_{V} \rho r_{i} (\partial \phi/\partial r_{i}) dV$, where $\phi$ is the gravitational potential.} it is necessary to evaluate each term in addition to the sign of the surface terms. Since the evaluation of one type of energy is rather independent, it is hardly possible to obtain from a single observational technique all the relevant quantities necessary to perform this comparison. Thus, in a first approximation, one might attempt to measure the individual energy ratios (thermal/gravity, magnetic/gravity, and kinetic/gravity) separately. This information is available to us in the simulations and can be used to evaluate the corrections that should be applied to the observed classical indicators in order to take into account the effect of the surface energy terms (which represent the effect of the environment). 

\section{Simulations}
In this study, we have analyzed a sample of 3D, isothermal, magnetized, driven, and self-gravitating MC simulations. The simulations vary by the strength of the initial magnetic field in the parent cloud. The MHD equations are solved using the TVD code described in Kim et al. (1999) at a numerical resolution of $256^{3}$. Periodic boundary conditions are used in the three directions. In physical units, the simulation box represents a scale of 4 pc, an average density of $\bar{n}=500$ cm$^{-3}$, a thermal sound speed of 0.2 km s$^{-1}$, and turbulence is driven such as to maintain a sonic Mach number of 10 (i.e., the ${\it rms}$ velocity is 2 km s$^{-1}$). The box has an initial Jeans number of $4$. Results are presented at epochs that precede the appearance of runaway gravitational collapse in any core in order to avoid violating the Truelove criterion (Truelove et al. 1997). CCs are identified in the simulations by using a clump-finding algorithm based on a density threshold and a friend-of-friend criteria. The algorithm is described in more detail in Dib et al. (2006a). The sample of objects selected in each simulation is a mix of CCs identified at 5 density thresholds (7.5, 15, 30, 60, and 100~$\bar{n}$). 

\begin{figure}
\plotone{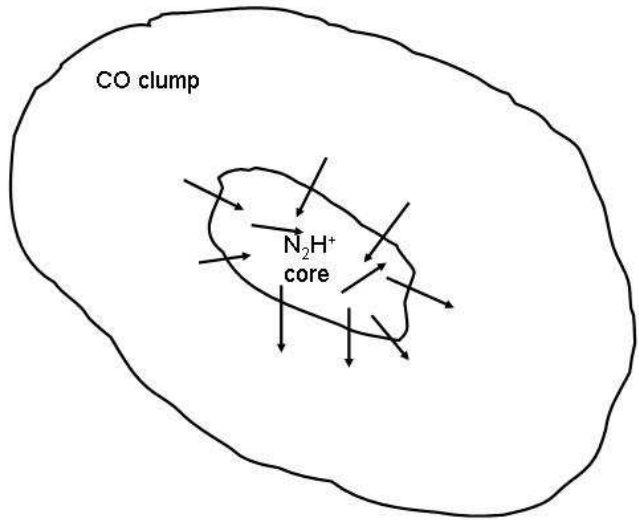}
\caption{A schematic view on how to measure the kinetic surface energy term for a dense molecular core using a non-depleted high density tracer such as N$_{2}$H$^{+}$. A second lower density tracer (e.g., a CO line) is needed to obtain information about the gas dynamics outside the N$_{2}$H$^{+}$ core boundary. The surface kinetic energy term is obtained by integrating the velocity gradients on the boundary.}
\label{fig2}
\end{figure}

\section{Results and Conclusions}
In the following, we show, as an example, the results for a model where the parent cloud in strongly supercritical, i.e., with an initial weak magnetic field of 4.6 $\mu$G and a mass-to magnetic flux ratio, normalized to the critical value for gravitational collapse, of $\mu_{box}=8.8$). In this model, we observe the existence of both collapsed, and other, non-collapsing, long-lived condensations (see Dib et al. 2006a for details). Fig.~\ref{fig1} (left) displays the relationship between the CCs ratios of thermal to gravitational energies $(E_{th}-\tau_{th})/|W|$ and their respective ($1/J_{c}^{2}$) estimates at the selected timestep. The two quantities are well correlated (stars in Fig.~\ref{fig1}). A linear fit is over-plotted to the data and is seen to follow:

\begin{equation}
\frac{(E_{th}-\tau_{th})} {|W|} =10^{-1.10 \pm 0.07}  \left (\frac {1}{J_{c}^{2}}\right)^{0.64 \pm 0.05}.
\end{equation} 

This relation provides the correction that ought to be applied to estimates of $J_{c}$ in CCs in order to get a more accurate estimate of the true thermal to gravitational energies ratio which takes into account the effect of the surface thermal energy term. If the surface terms are omitted ($\tau_{th}=0$, triangles in Figs.~\ref{fig1}), the linear fit shows, as expected, a slope close to 1 (0.99 $\pm$ 0.05). The comparison of the kinetic+thermal to gravitational energies ratios to $\alpha_{vir}$ is complicated by the fact that the quantity $(E_{k}+E_{th}-\tau_{k}-\tau_{th})$ can be of either sign. However, for both positive and negative families, Fig.~\ref{fig1} (right) do not show any correlation between $\alpha_{vir}$ and the (kinetic+thermal)/gravity energy ratios (If the surface terms are omitted the slope is close to 1, i.e., $0.90 \pm 0.09$, triangles). This result is essentially due to the complexity of the surface kinetic energy term, $\tau_{k}$, which varies substantially from one object to another even when the CCs have a similar volume kinetic energy. Therefore, it seems  likely that the evaluation of $\tau_{k}$ should be made directly from the observations. A simple method might be to integrate the kinetic energy for pixels which lie on or close to the boundary as defined arbitrarily by the molecular tracer and compare this quantity to the volume kinetic energy. A second method (Fig.~\ref{fig2}) might rely on measuring the velocity gradients across the boundary of the CC. Such method requires high spatial resolution in order to sample the boundary and the use of two distinct molecular tracers such that the line of sight velocity measurements at the inner side of the boundary would not be affected by material which lies outside the boundary. Making some additional assumptions such as the isotropy of turbulent motions in the CC, it might be possible to use the velocity gradients in order to calculate $\tau_{k}$ (the density at the boundary is equal to the critical density of the high density tracer). A detailed study, at higher numerical resolution, and including the comparison of the magnetic terms will be presented in a future work (Dib et al. 2006b in preparation)       

\acknowledgements 
S. D. is supported by a postdoctoral fellowship from the Korea Astronomy and Space Science Institute (KASI).

\end{document}